\documentclass[twocolumn]{aastex6}

\usepackage{amsmath}

\usepackage{enumitem}
\setitemize{itemsep=0pt}

\usepackage{listings}
\lstdefinestyle{codeintext}{language=Python, basicstyle=\ttfamily} 
\newcommand{\code}{\lstinline[style=codeintext]}

\newcommand{\hcode}{\large\normalfont\texttt} 

\begin{document}

\title{\MakeLowercase{\hcode{cluster-lensing}}: a \hcode{P}\MakeLowercase{\hcode{ython}} Package for Galaxy Clusters and Miscentering}

\author{Jes Ford\altaffilmark{1} and Jake VanderPlas\altaffilmark{2}}
\affil{eScience Institute, University of Washington \\ 3910 15th Ave NE, WRF Data Science Studio \\ Seattle, WA 98195-1570, USA}

\altaffiltext{1}{jesford@uw.edu}
\altaffiltext{2}{jakevdp@uw.edu}

\shorttitle{\hcode{cluster-lensing}}
\shortauthors{Ford \& VanderPlas}

\begin{abstract}
We describe a new open source package for calculating properties of galaxy clusters, including NFW halo profiles with and without the effects of cluster miscentering. This pure-\code{Python} package, \code{cluster-lensing}, provides well-documented and easy-to-use classes and functions for calculating cluster scaling relations, including mass-richness and mass-concentration relations from the literature, as well as the surface mass density $\Sigma(R)$ and differential surface mass density $\Delta\Sigma(R)$ profiles, probed by weak lensing magnification and shear. Galaxy cluster miscentering is especially a concern for stacked weak lensing shear studies of galaxy clusters, where offsets between the assumed and the true underlying matter distribution can lead to a significant bias in the mass estimates if not accounted for. This software has been developed and released in a public GitHub repository, and is licensed under the permissive MIT license. The \code{cluster-lensing} package is archived on Zenodo \citep{clusterlensing}. Full documentation, source code, and installation instructions are available at \url{http://jesford.github.io/cluster-lensing/}.
\end{abstract}


\setcounter{section}{0}
\setcounter{subsection}{0}
\setcounter{subsubsection}{0}


\section{Introduction}
\label{intro}

Clusters of galaxies are the largest gravitationally collapsed structures to have formed in the history of the universe.  As such, they are interesting both from a cosmological as well as an astrophysics perspective. In the former case, the galaxy cluster number density as a function of mass (the cluster mass function) is a probe of cosmological parameters including the fractional matter density $\Omega_{\rm m}$ and the normalization of the matter power spectrum $\sigma_8$. Astrophysically, the deep potential wells of galaxy clusters are environments useful for testing theories of general relativity, galaxy evolution, and gas and plasma physics, among other things \citep{Voit05}.

The common thread among these diverse investigations is the requisite knowledge of the mass of the galaxy cluster, which is largely composed of its invisible dark matter halo. Although many techniques exist for estimating the total mass of these systems, weak lensing has emerged as somewhat of a gold standard, since it is sensitive to the mass itself, and not to the dynamical state or other biased tracers of the underlying mass. Scaling relations between weak lensing derived masses, and other observables, including richness, X-ray luminosity and temperature, for examples, are typically calibrated from large surveys and extrapolated to clusters for which gravitational lensing measurements are impossible or unreliable. Since weak lensing masses are often considered the ``true'' masses, against which other estimates are compared \citep[e.g.][]{Leauthaud10, vonderLinden14, Hoekstra15}, it is paramount that cluster masses from weak lensing modeling are as unbiased as possible.

For stacked weak lensing measurements of galaxy clusters, an important source of bias in fitting a mass model is the inclusion of the effect of miscentering offsets. Miscentering occurs when the center of the mass distribution -- the dark matter halo -- does not perfectly coincide with the assumed center around which tangential shear (or magnification) profiles are being measured. Candidate centers for galaxy clusters are necessarily chosen from observational proxies, and often include a single galaxy, such as the brightest or most massive member, or the centroid of some extended quantity like the peak of X-ray emission or average of galaxy positions \citep{George12}. The particular choice of center may be offset from the true center due to interesting physical processes such as recent mergers and cluster evolution, or simply due to misidentification of the proxy of interest \citep{Johnston07}.

The miscentering effect on the stacked weak lensing profile can be included in a proper modeling of the measurement, as done in \citet{Johnston07, Mandelbaum10, Oguri11, George12, Sehgal13, Oguri14, Ford14, Ford15, Simet16}. The inclusion of this effect commonly assumes a form for the distribution of offsets, such as a Rayleigh distribution in radius (which represents a 2D Gaussian in the plane of the sky). This distribution is convolved with the standard (centered) halo profile to obtain the miscentered version. Software for calculating these miscentered weak lensing profiles was developed in order to produce results in \citet{Ford14, Ford15}, and has recently been publicly released to the astronomical community \citep{clusterlensing}.\footnote{\url{https://github.com/jesford/cluster-lensing}}

When many different gravitational lenses are stacked, as is often necessary to increase signal-to-noise for weak lensing measurements, care must be taken in the interpretation of the average signal. The issue here is that the (differential) surface mass density is not a linear function of the mass, so the average of many stacked profiles does not directly yield the average mass of the lens sample. Care must be taken to consider the underlying distribution of cluster masses as well as the redshifts of lenses and sources, all of which affect the amplitude of the measured lensing profile. One approach to this is to use a so-called composite-halo approach \citep[e.g.][]{Hildebrandt11, Ford12, Ford14, Ford15, Simet16}, where profiles are calculated for all individual lens objects and then averaged together to create a model that can be fit to the measurement. The \code{ClusterEnsemble()} class discussed in Section \ref{clusters} is designed with this approach in mind.

A popular model for the dark matter distribution in a gravitationally collapsed halo, such as a galaxy cluster, is the Navarro, Frenk, and White (NFW) model. This density profile (given in Equation \ref{nfw3d} below) was determined from numerical simulations that included the dissipationless collapse of density fluctuations under gravity \citep{nfw97}. The simpler Singular Isothermal Sphere density model, which only has one free parameter in contrast to the two for NFW, does not tend to fit the inner profiles of halos well and is also unphysical in that the total mass diverges \citep{Schneider06_WeakGravLens}. Other models such as the generalized-NFW and the Einasto profile tend to better describe the full radial distribution of dark matter in halos, at the expense of adding a third parameter to characterize the inner slope of the density profiles \citep[see e.g. discussion in][]{Dutton14}. In the software package presented in this work we only include the standard 2-parameter NFW model, but future work should make alternative models available as well.


\section{Description of the Code}
\label{code}

In this section we demonstrate each of the individual modules available in the \code{cluster-lensing} package. In Section \ref{nfw} we describe a class for calculating surface mass density profiles directly from NFW and cosmological parameters. Next we outline the functions available for mass-concentration relations in Section \ref{cofm}. Then in Section \ref{clusters} we present the class \code{ClusterEnsemble()}, and its related functions, which tie together the previously discussed functionality into a framework for easily manipulating and producing profiles for multiple galaxy clusters at once, from common observational quantities. Much of the content of this section comes directly from the online documentation.\footnote{\url{http://jesford.github.io/cluster-lensing/}} Throughout the modules, dimensionful quantities are labelled as such by means of the \code{astropy.units} package.


\subsection{\normalfont{\hcode{nfw.py}}}
\label{nfw}

The \code{nfw.py} module contains a single class called \code{SurfaceMassDensity()}, which computes the surface mass density $\Sigma(R)$ and the differential surface mass density $\Delta\Sigma(R)$ using the class methods \code{sigma_nfw()} and \code{deltasigma_nfw()}, respectively. These profiles are calculated according to the analytical formulas first derived by \citet{Wright00}, assuming the spherical NFW model, and can be applied to any dark matter halo: \emph{this module is not specific to galaxy clusters}.

The 3-dimensional density profile of an NFW halo is given by
\begin{equation}\label{nfw3d}
\rho(r) = \frac{\delta_{\rm c} \rho_{\rm crit}}{(r/r_{\rm s})(1+r/r_{\rm s})^2},
\end{equation}
where $r_{\rm s}$ is the cluster scale radius, $\delta_{\rm c}$ is the characteristic halo overdensity, and $\rho_{\rm crit} = \rho_{\rm crit}(z)$ is the critical energy density of the universe at the lens redshift. These three parameters\footnote{or, in the case of calculating multiple NFW halos at once, three array-like objects representing each of these parameters} must be specified when instantiating the class \code{SurfaceMassDensity()}, via the arguments \code{rs}, \code{delta_c}, and \code{rho_crit}, respectively. The units on \code{rs} are assumed to be Mpc, \code{delta_c} is dimensionless, and \code{rho_crit} is in $M_{\odot}{\rm Mpc}^{-1}{\rm pc}^{-2}$, although the actual inclusion of the \code{astropy.units} on these variables is optional. The user will probably also want to choose the radial bins for the calculation, which are specified via the keyword argument \code{rbins}, in Mpc. The surface mass density is the integral along the line-of-sight of the 3-dimensional density:
\begin{equation}
\Sigma(R) = 2 \int_0^{\infty} \rho(R,y) {\rm d}y.
\end{equation}
Here $R$ is the projected radial distance (in the plane of the sky).

We can adopt the dimensionless radius $x \equiv R/r_{\rm s}$ and, following from \citet{Wright00}, show that:
\begin{equation}\label{sigma}
\Sigma(x) = 2 r_{\rm s} \delta_{\rm c} \rho_{\rm crit} f(x),
\end{equation}
where $f(x) = $
\begin{equation}
    \begin{cases}
        \frac{1}{x^2 - 1} \left( 1 - \ln{ \left[ \frac{1}{x} + \sqrt{ \frac{1}{x^2} - 1} \right]} / \sqrt{1 - x^2} \right), & \text{for } x < 1; \\
        1/3, & \text{for } x = 1; \\
        \frac{1}{x^2 - 1} \left( 1 - \arccos{(1/x)} / \sqrt{x^2 - 1} \right), & \text{for } x > 1.
    \end{cases}
\end{equation}

The differential surface mass density probed by shear is calculated from the definition
\begin{equation}\label{dsigma}
\Delta\Sigma(x) \equiv  \overline{\Sigma}(<x) - \Sigma(x),
\end{equation}
where
\begin{equation}\label{sigbar}
\overline{\Sigma}(<x) = \frac{2}{x^2} \int_0^{x} \Sigma(x') x' {\rm d}x'.
\end{equation}

We can rewrite the differential surface mass density in the form in which it is computed in \code{nfw.py}:
\begin{equation}
     \Delta\Sigma(x) = r_{\rm s} \delta_{\rm c} \rho_{\rm crit} g(x),
\end{equation}
where  $g(x) = $
\begin{equation}
    \begin{cases}

    \left[ \frac{4 / x^2 + 2/ (x^2 - 1)}{\sqrt{1 - x^2}} \right] \ln \left( \frac{1 + \sqrt{(1-x) / (1+x)} }{1 - \sqrt{(1-x) / (1+x)} } \right) \\ + \frac{4}{x^2} \ln \frac{x}{2} - \frac{2}{(x^2 - 1)}, & \text{for } x < 1; \\

    (10/3) + 4 \ln(1/2), & \text{for } x = 1; \\

    \left[ \frac{8}{x^2 \sqrt{x^2 - 1}} + \frac{4}{(x^2 - 1)^{3/2}} \right] \arctan\sqrt{\frac{x-1}{1+x}} \\ + \frac{4}{x^2} \ln \frac{x}{2} - \frac{2}{(x^2 - 1)}, & \text{for } x > 1.

    \end{cases}
\end{equation}

Running \code{sigma_nfw()} or \code{deltasigma_nfw()}, with only a specification of halo properties \code{rs}, \code{delta_c}, \code{rho_crit}, and radial bins \code{rbins}, will lead to the calculation of halo profiles according to Equations \ref{sigma} and \ref{dsigma} outlined above.

\begin{lstlisting}
from clusterlensing import SurfaceMassDensity
rbins = [0.1, 0.5, 1.0, 2.0, 4.0] # Mpc
smd = SurfaceMassDensity(rs=[0.1],
                         rho_crit=[0.2],
                         delta_c=[9700.],
                         rbins=rbins)
sigma = smd.sigma_nfw()
# surface mass density with default units
sigma[0]
<Quantity [ 129.33333333, 11.64751032, 3.33992059,
0.89839601, 0.23327149] solMass / pc2>

# surface mass density with no units
sigma[0].value
array([ 129.33333333, 11.64751032, 3.33992059,
0.89839601, 0.23327149])
\end{lstlisting}

These are the standard centered NFW profiles, under the assumption that the peak of the halo density distribution perfectly coincides with the identified halo center. This may not be a good assumption, however, and the user can instead run these calculations for miscentered halos by specifying the optional input parameter \code{offsets}. This parameter sets the width of a distribution of centroid offsets, assuming a 2-dimensional Gaussian distribution on the sky. This offset distribution is equivalent to, and implemented in code as, a uniform distribution in angle and a Rayleigh probability distribution in radius:
\begin{equation}\label{PofR}
P(R_{\mathrm{off}})=\frac{R_{\mathrm{off}}}{\sigma_{\mathrm{off}}^2}\ \mathrm{exp}\bigg[-\frac{1}{2}\bigg(\frac{R_{\mathrm{off}}}{\sigma_{\mathrm{off}}}\bigg)^2\ \bigg].
\end{equation}
The parameter \code{offsets} is equivalent to $\sigma_{\rm off}$ in this equation.

\begin{lstlisting}
from clusterlensing import SurfaceMassDensity
rbins = [0.1, 0.5, 1.0, 2.0, 4.0]
# single miscentered halo profile
smd = SurfaceMassDensity(rs=[0.1],
                         rho_crit=[0.2],
                         delta_c=[9700.],
                         rbins=rbins,
                         offsets=[0.3])
sigma = smd.sigma_nfw()
sigma[0]
<Quantity [ 38.60655298, 17.57285034, 4.11253461,
0.93809627, 0.23574031] solMass / pc2>

# example calculating multiple profiles
smd = SurfaceMassDensity(rs=[0.1,0.2,0.2],
                         rho_crit=[0.2,0.2,0.2],
                         delta_c=[9700,9700,9000],
                         rbins=rbins,
                         offsets=[0.3,0.3,0.3])
sigma = smd.sigma_nfw()
sigma
<Quantity [[  38.60655298,  17.57285034,   4.11253461,   0.93809627,   0.23574031],
           [ 181.91820855,  92.86651598,  27.34020647,   6.94677803,  1.81488253],
           [ 168.79009041,  86.16480864,  25.36720188,   6.44546415,  1.68391163]] solMass / pc2>
\end{lstlisting}

The miscentered surface mass density profiles are given by the centered profiles (Equations \ref{sigma} and \ref{dsigma}), convolved with the offset distribution (Equation \ref{PofR}). We follow the offset halo formalism first written down by \citet{Yang06}, and applied to cluster miscentering by, e.g. \citet{Johnston07, George12, Ford14, Ford15, Simet16}. Specifically, we calculate the offset surface mass density $\Sigma^{\rm off}$ as follows:
\begin{equation}\label{sigma_off}
\Sigma^{\rm off}(R) = \int_{0}^{\infty} \Sigma(R | R_{\rm off})\ P(R_{\rm off})\ {\rm d}R_{\rm off}
\end{equation}
\begin{equation}\label{sigma_RgivenRoff}
\Sigma(R|R_{\mathrm{off}})=\frac{1}{2\pi}\int_{0}^{2\pi}\Sigma(r) \mathrm{d}\theta
\end{equation}
Here $r = \sqrt{R^2+R_{\mathrm{off}}^2-2RR_{\mathrm{off}}\cos(\theta)}$ and $\theta$ is the azimuthal angle \citep{Yang06}. The $\Delta\Sigma^{\rm off}$ profile is calculated from $\Sigma^{\rm off}$, in analogy with Equations \ref{dsigma} and \ref{sigbar}.

\begin{lstlisting}
from clusterlensing import SurfaceMassDensity
rbins = [0.1, 0.5, 1.0, 2.0, 4.0]
# perfectly centered DeltaSigma profile
smd = SurfaceMassDensity(rs=[0.1],
                         rho_crit=[0.2],
                         delta_c=[9700.],
                         rbins=rbins)
deltasigma = smd.deltasigma_nfw()
deltasigma[0]
<Quantity [ 108.78445455, 25.47093418, 10.29627483,
3.71631903, 1.23840727] solMass / pc2>

# miscentered DeltaSigma profile
smd = SurfaceMassDensity(rs=[0.1],
                         rho_crit=[0.2],
                         delta_c=[9700.],
                         rbins=rbins,
                         offsets=[0.3])
deltasigma = smd.deltasigma_nfw()
deltasigma[0]
<Quantity [ 0.71370144, 9.35821817, 8.90118561,
3.6475417, 1.23610325] solMass / pc2>
\end{lstlisting}


\subsection{\hcode{cofm.py}}
\label{cofm}

The \code{cofm.py} module currently contains three functions, each of which calculates halo concentration from mass, redshift, and cosmology, according to a prescription given in the literature. These functions are \code{c_DuttonMaccio()} \citep[for calculations following][]{Dutton14}, \code{c_Duffy()} \citep[following][]{Duffy08}, and \code{c_Prada()} \citep[for][]{Prada12}. Halo mass-concentration relations are an area of active research, and there have been discrepancies between results from different observations and simulations, and disagreement surrounding the best choice of model \citep[see e.g.][]{Dutton14, Klypin16}. We do not aim to join this discussion here, but focus on outlining the functionality provided by the \code{cluster-lensing} package, for calculating these different concentration values.

All three functions require two input parameters (scalars or array-like inputs), which are the halo redshift(s) \code{z} and the halo mass(es) \code{m}. Specifically, the latter is assumed to correspond to the $M_{200}$ mass definition, in units of solar masses. $M_{200}$ is the mass interior to a sphere of radius $r_{200}$, within which the average density is $200\rho_{\rm crit}(z)$.

The default cosmology used is from the measurements by the \citet{PlanckXVI}, which is imported from the module \code{astropy.cosmology.Planck13}. However, the user can specify alternative cosmological parameters. For calculating concentration according to either the \citet{Duffy08} or the \citet{Dutton14} prescription, the only cosmological parameter required is the Hubble parameter, which can be passed into \code{c_Duffy()} or \code{c_DuttonMaccio()} as the keyword argument \code{h}. For the \citet{Prada12} concentration, the user would want to specify \code{Om_M} and \code{Om_L} (the fractional energy densities of matter and the cosmological constant) in addition to \code{h}, in the call to \code{c_Prada()}.

The \code{c_DuttonMaccio()} calculation of concentration is done according to the power-law
\begin{equation}
\log_{10} c_{200} = a + b \log_{10}(M_{200} / [10^{12} h^{-1} M_{\odot}]),
\end{equation}
where
\begin{equation}
a = 0.52 + 0.385\ {\rm exp}[-0.617\ z^{1.21}],
\end{equation}
\begin{equation}
b = -0.101 + 0.206 z.
\end{equation}
The above three equations map to Equations 7, 11, and 10, respectively in \citet{Dutton14}. The values in these expressions were determined from simulations of halos between $0 < z < 5$, spanning over 5 orders of magnitude in mass, and were shown to match observational measurements of low-redshift galaxies and clusters \citep{Dutton14}. This concentration-mass relation is the default one used by the \code{clusters.py} module, discussed in Section \ref{clusters}.

\begin{lstlisting}
from clusterlensing import cofm
# single 10**14 Msun halo at z=1
cofm.c_DuttonMaccio(0.1, 1e14)
array([ 5.13397936])
# example with multiple halos
cofm.c_DuttonMaccio([0.1, 0.5], [1e14, 1e15])
array([ 5.13397936,  3.67907305])
\end{lstlisting}

The concentration calculation in \code{c_Duffy()} is
\begin{equation}\label{c_Duffy}
c_{200} = A\ (M_{200} / M_{\rm pivot})^B \ (1 + z)^C,
\end{equation}
where
\begin{equation}
\{A, B, C\} = \{5.71, -0.084, -0.47\},
\end{equation}
\begin{equation}
M_{\rm pivot} = 2 \times 10^{12}\ h^{-1} M_{\odot}.
\end{equation}
Equation \ref{c_Duffy} above corresponds to Equation 4 in \citet{Duffy08}. The values for $A$, $B$, and $C$ can be found in Table 1 of that work, where they are specific to the ``full'' (relaxed and unrelaxed) sample of simulated NFW halos, spanning the redshift range $0 < z < 2$. $M_{\rm pivot}$ can be found in the caption of their Table 1 as well. One caveat with this relation is that the cosmology used in creating the \citet{Duffy08} simulations was that of the now outdated WMAP5 experiment \citep{WMAP5}.

\begin{lstlisting}
from clusterlensing import cofm
# default cosmology (h=0.6777)
cofm.c_Duffy([0.1, 0.5], [1e14, 1e15])
array([ 4.06126115, 2.89302767])
# with h=1
cofm.c_Duffy([0.1, 0.5], [1e14, 1e15], h=1)
array([ 3.93068341, 2.80001099])
\end{lstlisting}

The \code{c_Prada()} concentration calculation is much more complex, and written in terms of $\sigma(M_{200}, x_{\rm p})$, the rms fluctuation of the density field. The \citet{Prada12} halo concentration is given by\footnote{we use the subscript ``p'' to distinguish some variables in the equations from \citet{Prada12} from those in the current work}
\begin{equation}\label{c_Prada}
\begin{split}
c_{200} = 2.881 B_0 (x_{\rm p}) \bigg[ \left( \frac{B_1(x_{\rm p}) \sigma(M_{200},x_{\rm p})}{1.257} \right)^{1.022} + 1 \bigg] \\
\times\ {\rm exp}\left( \frac{0.06}{[B_1(x_{\rm p}) \sigma(M_{200},x_{\rm p}) ]^2} \right).
\end{split}
\end{equation}
The cosmology and redshift dependence is encoded by the variable $x_{\rm p}$, which is
\begin{equation}
x_{\rm p} = \left( \frac{\Omega_{\Lambda, 0}}{\Omega_{{\rm m}, 0}} \right)^{1/3} (1 + z)^{-1}.
\end{equation}
The functions within Equation \ref{c_Prada} are as follows:
\begin{equation}
\sigma(M_{200},x_{\rm p}) = D(x_{\rm p}) \frac{16.9 y_{\rm p}^{0.41}}{1 + 1.102 y_{\rm p}^{0.2} + 6.22 y_{\rm p}^{0.333}}
\end{equation}
\begin{equation}
y_{\rm p} \equiv \frac{10^{12} h^{-1} M_{\odot}}{M_{200}}
\end{equation}
\begin{equation}
D(x_{\rm p}) = \frac{5}{2} \left( \frac{\Omega_{{\rm m}, 0}}{\Omega_{\Lambda, 0}} \right)^{1/3} \frac{\sqrt{1 + x_{\rm p}^3}}{x_{\rm p}^{3/2}} \int_0^{x_{\rm p}} \frac{x^{3/2} {\rm d}x}{(1 + x^3)^{3/2}}
\end{equation}
\begin{equation}
B_0(x_{\rm p}) = \frac{c_{\rm min}(x_{\rm p})}{c_{\rm min}(1.393)}
\end{equation}
\begin{equation}
B_1(x_{\rm p}) = \frac{\sigma^{-1}_{\rm min}(x_{\rm p})}{\sigma^{-1}_{\rm min}(1.393)}
\end{equation}
\begin{equation}
c_{\rm min}(x_{\rm p}) = 3.681 + 1.352 \bigg[ \frac{1}{\pi} \arctan[6.948 (x_{\rm p} - 0.424)] + \frac{1}{2} \bigg]
\end{equation}
\begin{equation}
\sigma^{-1}_{\rm min}(x_{\rm p}) = 1.047 + 0.599 \bigg[ \frac{1}{\pi} \arctan[7.386 (x_{\rm p} - 0.526)] + \frac{1}{2} \bigg]
\end{equation}
In order of appearance above, beginning with our Equation \ref{c_Prada}, these equations correspond to Equations 14-17, 13, 23a, 23b, 12, 18a, 18b, 19, 20 in \citet{Prada12}. The numerical values in these equations were obtained empirically from the simulations described in that work.
\begin{lstlisting}
from clusterlensing import cofm
cofm.c_Prada([0.1, 0.5], [1e14, 1e15])
array([ 5.06733941,  5.99897362])
cofm.c_Prada([0.1, 0.1, 0.1], [1e13, 1e14, 1e15])
array([ 5.71130928, 5.06733941, 5.30163572])
\end{lstlisting}
The last code example demonstrates the controversial feature of the \citet{Prada12} mass-concentration relation -- an upturn in concentration values for the highest mass halos. This is in opposition to the canonical view that higher mass halos have lower concentrations \citep{nfw96, nfw97, Jing00, Bullock01}.


\subsection{\normalfont{\hcode{clusters.py}}}
\label{clusters}
The \code{clusters.py} module is designed to provide a catalog-level tool for calculating, tracking, and updating galaxy cluster properties and profiles, through structuring data from multiple clusters as an updatable Pandas Dataframe, and providing an intelligent interface to the other modules discussed in Sections \ref{nfw} and \ref{cofm}. This module contains a single class \code{ClusterEnsemble()}, as well as three functions, \code{mass_to_richness()}, \code{richness_to_mass()}, and \code{calc_delta_c()}.

The function \code{calc_delta_c()} takes a single input parameter, the cluster concentration \code{c200} (e.g. as calculated by one of the functions in \code{cofm.py}), and returns the characteristic halo overdensity:
\begin{equation}
\delta_{\rm c} = \left( \frac{200}{3} \right) \frac{c_{200}^3}{\ln(1 + c_{200}) - c_{200}/(1 + c_{200})}.
\end{equation}
Both input and output are dimensionless here. For example, to convert a concentration value of $c_{200} = 5$ to $\delta_{\rm c}$, you could do:
\begin{lstlisting}
from clusterlensing.clusters import calc_delta_c
calc_delta_c(5)
8694.8101906193315
\end{lstlisting}

The pair of functions \code{mass_to_richness()} and \code{richness_to_mass()}, as their names imply, perform conversions between cluster mass and richness. The only required input parameter to \code{mass_to_richness()} is the \code{mass}, and likewise the only required input to \code{richness_to_mass()} is \code{richness}. The calculations assume a power-law form for the relationship between these variables:
\begin{equation}\label{massrich}
M_{200} = M_0 \left( \frac{N_{200}}{20} \right) ^ \beta.
\end{equation}
Here $M_0$ is the normalization, which defaults to $2.7 \times 10^{13}$, but can be changed in the call to either function by setting the \code{norm} keyword argument. The power-law slope $\beta = 1.4$ by default, but can be set by specifying the optional \code{slope} input parameter. When these functions are invoked by the \code{ClusterEnsemble()} class, they are applied to the particular mass definition $M_{200}$, and assume units of $M_{\odot}$. However the functions themselves do not assume a mass definition or unit, and can be generalized to any parameter (or type of richness) that has a power-law relationship with mass.

\begin{lstlisting}
from clusterlensing.clusters import \
mass_to_richness, richness_to_mass
richness_to_mass(50)
97382243648736.9
mass_to_richness(97382243648736.9)
50.0
# specify other power-law parameters
richness_to_mass(20, slope=1.5, norm=1e14)
100000000000000.0
\end{lstlisting}

The \code{ClusterEnsemble()} class creates, modifies and tracks a Pandas DataFrame containing the properties and attributes of many galaxy clusters at once. When given a new or updated cluster property, it calculates and updates all dependent cluster properties, treating each cluster (row) in the DataFrame as an individual object. This makes it easy to calculate the $\Sigma(R)$ and $\Delta\Sigma(R)$ weak lensing profiles for many different mass clusters at different redshifts, with a single command. In contrast to using the \code{SurfaceMassDensity()} class discussed in Section \ref{nfw}, the user only needs to specify the cluster redshifts and either of the mass or richness. If richness is supplied, then mass is calculated from it, assuming the form of Equation \ref{massrich} (which is customizable); if mass is specified instead, than the inverse relation is used to calculate richness. In either case the changes are propagated to any dependent variables.

\begin{lstlisting}
from clusterlensing import ClusterEnsemble
z = [0.1,0.2,0.3]
c = ClusterEnsemble(z)
n200 = [20, 20, 20]
c.n200 = n200

# display cluster dataframe
c.dataframe
     z  n200          m200      r200      c200       delta_c        rs
0  0.1    20  2.700000e+13  0.612222  5.839934  12421.201995  0.104834
1  0.2    20  2.700000e+13  0.591082  5.644512  11480.644557  0.104718
2  0.3    20  2.700000e+13  0.569474  5.442457  10555.781440  0.104636

# specify mass directly
c.m200 = [1e13, 1e14, 1e15]
c.dataframe
     z        n200          m200      r200      c200       delta_c        rs
0  0.1    9.838141  1.000000e+13  0.439664  6.439529  15599.114356  0.068276
1  0.2   50.956400  1.000000e+14  0.914520  4.979102   8612.362538  0.183672
2  0.3  263.927382  1.000000e+15  1.898248  3.886853   4947.982895  0.488377
\end{lstlisting}

The above examples also demonstrate that cluster masses are converted to concentrations and to characteristic halo overdensities. This assumes the default mass-concentration relation of the \code{c_DuttonMaccio()} form, or the user can instead specify another of the relations by setting the keyword \code{cm="Prada"} or \code{cm="Duffy"}, when the \code{ClusterEnsemble()} object is instantiated. Cosmology can also be specified upon instantiation, by setting the \code{cosmology} keyword to be any \code{astropy.cosmology} object that has an \code{h} and a \code{Om0} attribute. If not specified explicitly, the default cosmological model used is \code{astropy.cosmology.Planck13}. Here is an example of creating a \code{ClusterEnsemble()} object that uses the WMAP5 cosmology \citep{WMAP5} and the \citet{Duffy08} concentration:
\begin{lstlisting}
from astropy.cosmology import WMAP5 as cosmo
c = ClusterEnsemble(z, cm="Duffy",
                    cosmology=cosmo)
c.n200 = [20, 30, 40]
c.dataframe
     z  n200          m200      r200      c200      delta_c        rs
0  0.1    20  2.700000e+13  0.599910  4.520029  6920.955951  0.132723
1  0.2    30  4.763120e+13  0.702040  4.136873  5678.897592  0.169703
2  0.3    40  7.125343e+13  0.775889  3.851601  4849.836498  0.201446
\end{lstlisting}
Instead of using the \code{dataframe} attribute, which retrieves the Pandas DataFrame object itself, it might be useful to use the \code{show()} method, which prints additional information to the screen, including assumptions of the mass-richness relation:
\begin{lstlisting}
c.show()

Cluster Ensemble:

     z  n200          m200      r200      c200      delta_c        rs

0  0.1    20  2.700000e+13  0.599910  4.520029  6920.955951  0.132723
1  0.2    30  4.763120e+13  0.702040  4.136873  5678.897592  0.169703
2  0.3    40  7.125343e+13  0.775889  3.851601  4849.836498  0.201446

Mass-Richness Power Law:
M200 = norm * (N200 / 20) ^ slope
   norm: 2.7e+13 solMass
   slope: 1.4

# update the mass-richness parameters
# and show the resulting table
c.massrich_norm = 3e13
c.massrich_slope = 1.5
c.show()

Cluster Ensemble:
     z  n200          m200      r200      c200      delta_c        rs
0  0.1    20  3.000000e+13  0.621353  4.480202  6784.805438  0.138689
1  0.2    30  5.511352e+13  0.737028  4.086481  5526.615129  0.180358
2  0.3    40  8.485281e+13  0.822406  3.795500  4696.109606  0.216679

Mass-Richness Power Law:
M200 = norm * (N200 / 20) ^ slope
   norm: 3e+13 solMass
   slope: 1.5

\end{lstlisting}
The last example also demonstrates how the slope or normalization of the mass-richness relation can be altered, and the changes propagate from richness through to mass and other variables.

Then all the ingredients are in place to calculate halo profiles by invoking the \code{calc_nfw()} method, which interfaces to the \code{sigma_nfw()} and \code{deltasigma_nfw()} methods of the \code{SurfaceMassDensity()} class, and passes it the required inputs $\{ r_{\rm s}, \rho_{\rm crit}, \delta_{\rm c} \}$ for all the clusters behind the scenes. The value of $\rho_{\rm crit}$ is calculated at every cluster redshift using the (default \code{astropy.cosmology.Planck13} or user-specified) cosmological model. The user must specify the desired radial bins \code{rbins} in Mpc.
\begin{lstlisting}
import numpy as np
# create some logarithmic bins:
rmin, rmax = 0.1, 5.  # Mpc
rbins = np.logspace(np.log10(rmin),
                    np.log10(rmax),
                    num = 8)
# calculate the profiles:
c.calc_nfw(rbins=rbins)
# profiles now exist as attributes:
c.sigma_nfw
<Quantity [[ 128.97156123, 62.58323349,
27.01073105, 10.60607722, 3.88999449,
1.36360964, 0.46464366, 0.15563814],
[ 132.13989867, 64.10484454, 27.66159293,
10.85990257, 3.98265113, 1.39599118,
0.47565695, 0.15932308], [ 135.62272115,
65.782882, 28.38138702, 11.14121765,
4.08549675, 1.43196834, 0.48790043,
0.16342108]] solMass / pc2>
c.deltasigma_nfw
<Quantity [[ 105.3190568 , 72.43842908,
43.74538085, 23.44005481, 11.37085955,
5.10385452, 2.16011364, 0.87479771],
[ 107.98098357, 74.25022426, 44.82825347,
24.01505305, 11.64776118, 5.22744541,
2.21219956, 0.89582394], [ 110.88173507,
76.23087398, 46.01581348, 24.64741078,
11.95297965, 5.36391529, 2.26978998,
0.91909571]] solMass / pc2>
\end{lstlisting}

Similar to the direct use of \code{SurfaceMassDensity()}, discussed in Section \ref{nfw}, the miscentered profiles can be calculated from the \code{calc_nfw()} method, by supplying the optional \code{offsets} keyword with an array-like object of length equal to the number of clusters, where each element is the width of the offset distribution in Mpc ($\sigma_{\rm off}$ in Equation \ref{PofR}).
\begin{lstlisting}
c.calc_nfw(rbins=rbins, offsets=[0.3,0.3,0.3])
# the offset sigma profile is now:
c.sigma_nfw
<Quantity [[ 42.50844685, 39.74291121,
32.29894213, 18.50988719, 6.16284894,
1.89335218, 0.62609991, 0.20840423],
[ 68.10228964, 63.87901872, 52.56539317,
31.20890672, 11.17821854, 3.5884285,
1.20745376, 0.40574057], [ 95.16077234,
89.48298631, 74.29328561, 45.24074628,
17.06333763, 5.66481165, 1.93408383,
0.65518747]] solMass / pc2>
\end{lstlisting}

Although \code{SurfaceMassDensity()} from the \code{nfw.py} module, and \code{ClusterEnsemble().calc_nfw()} from the \code{clusters.py} module, are both capable of computing the same $\Sigma(R)$ and $\Delta\Sigma(R)$ profiles, each require different forms of input which would make sense for different use cases. For the studies in \citet{Ford15}, \citet{Ford14}, and \citet{Ford12}, the authors wanted do the profile computations for many clusters at once, while varying the mass and the miscentering offset distribution during the process of fitting the model to the data. What was known were the redshifts and mass proxies (cluster richness in \citealt{Ford15} and \citealt{Ford14}, and a previous mass estimate in \citealt{Ford12}), and mass-concentration relations from the literature, so the \code{ClusterEnsemble()} framework made sense. However, if someone wanted to simply calculate the NFW profiles according to the \citet{Wright00} formulation, then they might prefer to use \code{SurfaceMassDensity()} as a tool to get profiles directly from the NFW and cosmological parameters $r_{\rm s}$, $\delta_{\rm c}$, and $\rho_{\rm crit}(z)$.


\section{Example}
\label{ex}
As an example use case, we take the Canada-France-Hawaii Telescope Lensing Survey \citep[CFHTLenS;][]{Heymans12, Erben13} public galaxy cluster catalog, which is available on Zenodo\footnote{\url{http://dx.doi.org/10.5281/zenodo.51291}} \citep{3DMFcatalog}. This dataset was previously explored using a pre-release version of the \code{cluster-lensing} software in \citet{Ford14, Ford15}. The W1 field of this survey contains 10,745 galaxy cluster candidates in the redshift range $0.2 \le z \le 0.9$:

\begin{lstlisting}
import numpy as np
data = np.loadtxt("Clusters_W1.dat")
data.shape
(10745, 5)
data[0:4, :]  # print first 4 clusters
array([[ 34.8023, -7.01005, 0.3, 4.435, 10.],
       [ 34.9425, -7.38996, 0.5, 4.545, 21.],
       [ 34.8651, -6.69449, 0.5, 3.858, 6.],
       [ 34.6224, -7.32768, 0.5, 3.619, 8.]])
redshift = data[:, 2]
sig = data[:, 3]
richness = data[:, 4]
\end{lstlisting}

We select a subset of the lower redshift clusters that were detected at high significance. Then we import \code{clusterlensing} to create a dataframe of the cluster properties, of which we just print the first several, and calculate the NFW profiles.

\begin{lstlisting}

# select a subset
here = (sig > 15) & (redshift < 0.5)
sig[here].shape
(15,)
z = redshift[here]
n200 = richness[here]

import clusterlensing
c = clusterlensing.ClusterEnsemble(z)
c.n200 = n200
c.dataframe.head()
     z  n200          m200      r200      c200      delta_c        rs
0  0.4   181  5.897552e+14  1.531367  3.966101  5173.016417  0.386114
1  0.3   420  1.916332e+15  2.357815  3.658237  4332.615805  0.644522
2  0.4   176  5.670737e+14  1.511478  3.980218  5213.746469  0.379747
3  0.3   113  3.049521e+14  1.277703  4.341779  6324.420397  0.294281
4  0.4   162  5.049435e+14  1.454129  4.022285  5336.272412  0.361518

rbins = np.logspace(np.log10(0.1),
                    np.log10(10.0), num=20)
c.calc_nfw(rbins)
\end{lstlisting}

\begin{figure}
\figurenum{1}
\epsscale{1.2}
\plotone{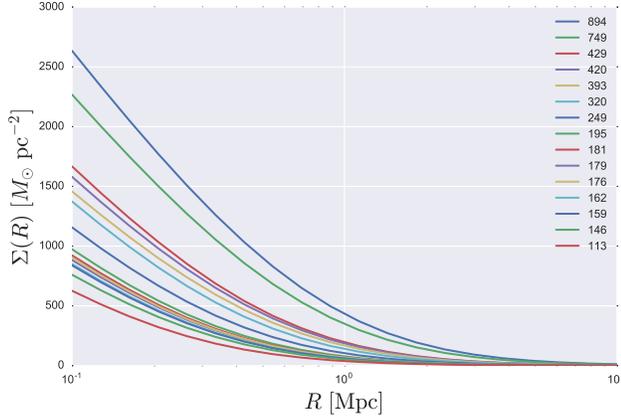}
\caption{Surface mass density profiles $\Sigma(R)$ for all 15 clusters used in this example. These are the most significant (S/N $>15$) clusters detected at low redshifts ($z < 0.5$) in the W1 field of CFHTLenS. See the text for links to download this public dataset. The legend gives the richness values estimated in \citet{Ford15} corresponding to each of these clusters, which are assumed to scale with mass. They are listed from highest to lowest richness, in the same order as the curves.}
\label{f1}
\end{figure}

Next we import the \code{matplotlib} and \code{seaborn} libraries and configure some settings to make our plots more readable. The first plot we create with the commands below presents the $\Sigma(R)$ profiles for every one of these 15 clusters, and is given in Figure \ref{f1}.

\begin{lstlisting}

import matplotlib.pyplot as plt
import seaborn as sns; sns.set()
import matplotlib
matplotlib.rcParams["axes.labelsize"] = 20
matplotlib.rcParams["legend.fontsize"] = 20

# strings for plots
raxis = "$R\ [\mathrm{Mpc}]$"
sgma = "$\Sigma(R)$"
sgmaoff = "$\Sigma^\mathrm{off}(R)$"
delta = "$\Delta$"
sgmaunits = " $[M_{\odot}\ \mathrm{pc}^{-2}]$"

# order from high to low richness
order = c.n200.argsort()[::-1]

for s, n in zip(c.sigma_nfw[order], c.n200[order]):
    plt.plot(rbins, s, label=str(int(n)))
plt.xscale("log")
plt.legend(fontsize=10)
plt.ylabel(sgma+sgmaunits)
plt.xlabel(raxis)
plt.tight_layout()
plt.savefig("f1.eps")
plt.close()  # output is Figure 1

\end{lstlisting}

\begin{figure}
\figurenum{2}
\epsscale{1.2}
\plotone{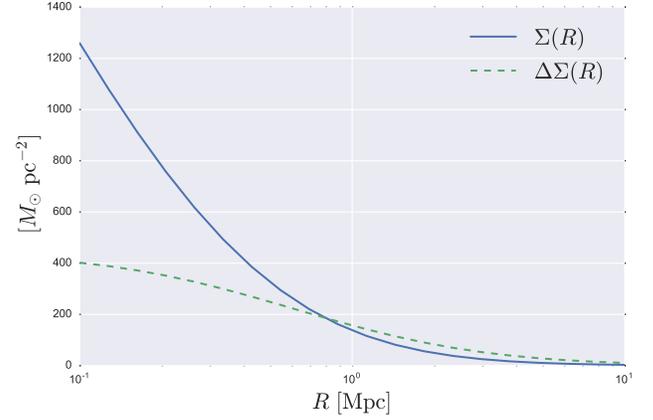}
\caption{The average of all the surface mass density profiles $\Sigma(R)$ for each of the clusters shown in Figure \ref{f1} is given in blue. The green curve is the average of all the individual differential mass density profiles $\Delta\Sigma(R)$. These curves assume clusters are perfectly centered on their NFW halos.}
\label{f2}
\end{figure}

If we had made a stacked measurement of the shear or magnification profile of these clusters, then we would want to know what the average profile of the stack looks like. Since we already have the individual profiles, we just need to calculate the mean across the 0$^{\rm th}$ axis of the \code{sigma_nfw} and \code{deltasigma_nfw} attribute arrays. The plot of these average profiles is given in Figure \ref{f2}.

\begin{lstlisting}

sigma = c.sigma_nfw.mean(axis=0)
dsigma = c.deltasigma_nfw.mean(axis=0)

plt.plot(rbins, sigma, label=sgma)
plt.plot(rbins, dsigma, "--", label=delta+sgma)
plt.legend()
plt.ylim([0., 1400.])
plt.xscale("log")
plt.xlabel(raxis)
plt.ylabel(sgmaunits)
plt.tight_layout()
plt.savefig("f2.eps")
plt.close()  # output is Figure 2

\end{lstlisting}

\begin{figure}
\figurenum{3}
\epsscale{1.2}
\plotone{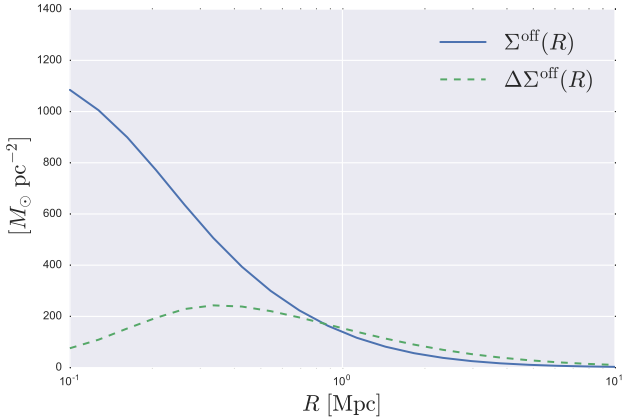}
\caption{Same as Figure \ref{f2} but for miscentered profiles. Cluster centroid offsets are assumed to follow a Rayleigh probability distribution (Equation \ref{PofR}, discussed in Section \ref{nfw}), which is convolved with the perfectly centered profiles to achieve this result.}
\label{f3}
\end{figure}

Finally, we may want to investigate whether cluster miscentering has a significant effect on our sample. We would calculate the miscentered profiles, given in Figure \ref{f3}, which could be compared to the centered profiles in Figure \ref{f2} to see which is a better fit to our measurement. Below we will assume that the miscentering offset distribution peaks at 0.1 Mpc.

\begin{lstlisting}

offsets = np.ones(c.z.shape[0]) * 0.1
c.calc_nfw(rbins, offsets=offsets)
sigma_offset = c.sigma_nfw.mean(axis=0)
dsigma_offset = c.deltasigma_nfw.mean(axis=0)

plt.plot(rbins, sigma_offset, label=sgmaoff)
plt.plot(rbins, dsigma_offset, "--",
         label=delta+sgmaoff)
plt.legend()
plt.ylim([0., 1400.])
plt.xscale("log")
plt.xlabel(raxis)
plt.ylabel(sgmaunits)
plt.tight_layout()
plt.savefig("f3.eps")
plt.close()  # output is Figure 3

\end{lstlisting}

The above example shows a simple application of \code{cluster-lensing} to a real dataset -- a subset of the CFHTLenS cluster catalog.  For this example we kept the customizations to a minimum, but as shown in Sections \ref{clusters}, the user can alter the parameters in the power-law relation used to convert richness to mass, choose the form of the mass-concentration relation assumed for the NFW profiles, and specify a particular background cosmology. When fitting a model produced by \code{cluster-lensing} to a measurement, one could iterate through parameters in this space by setting various attributes of the \code{ClusterEnsemble()} object \citep[as done, e.g. in][]{Ford14, Ford15}.


\section{Relation to Existing Code}

The \code{cluster-lensing} project offers some unique capabilities over other publicly-available software, most notably the cluster miscentering calculations. Here we attempt to compare the software presented in this work with other open source tools that we are aware of, and show how \code{cluster-lensing} fits into the larger ecosystem of astronomical software.

\code{Colossus} is a Python package aimed at cosmology, halo, and large-scale structure calculations \citep{ColossusCode}. It was used in work by \citet{Diemer15} and is made available under the MIT license\footnote{\url{http://www.benediktdiemer.com/code/}}. Much of the functionality of \code{cluster-lensing} appears to overlap with \code{Colossus}, including mass-concentration relations (although \code{Colossus} has the advantage of containing many more relations from the literature) and NFW surface mass density profiles. However, \code{cluster-lensing} also provides the miscentered halo calculations, which are are lacking from \code{Colossus}.

While \citet{ColossusCode} has chosen to implement basic cosmological calculations from scratch, \code{cluster-lensing} instead relies on external modules supplied by \code{astropy}.  The only dependencies claimed by \code{Colossus} are \code{numpy} \citep{NumPy} and \code{scipy} \citep{SciPy}, whereas \code{cluster-lensing} additionally requires \code{astropy} \citep{astropy13} and \code{pandas} \citep{Pandas}. Fewer dependencies might be seen as a positive feature of \code{Colossus}; on the other hand, \code{astropy} could be viewed as possibly a more robust source for standard astronomical and cosmological calculations, since it is maintained by a large community of developers.

Another related set of code is provided by J\"{o}rg Dietrich's NFW routines, archived on Zenodo \citep{DietrichNFW}, and available on GitHub\footnote{\url{https://github.com/joergdietrich/NFW}}. These \code{Python} modules calculate NFW profiles for $\Sigma(R)$ and $\Delta\Sigma(R)$, as well as the 3-dimensional density profiles and total mass and projected mass inside a given radius. \code{cluster-lensing} goes beyond the functionality of \citet{DietrichNFW} by supplying means for calculating cluster miscentering, and having a built-in framework for handling many halos at once. For \citet{DietrichNFW}, the user must provide the halo concentration (along with mass and redshift) to the \code{NFW()} class, but additional routines are available for converting mass to concentration, including \citet{Duffy08} and another by \citet{Dolag04} (a partial overlap with the mass-concentration relations provided by \code{cluster-lensing}). \citet{DietrichNFW} depends on \code{astropy} for cosmological calculations and units, similar to \code{cluster-lensing}, as well as the \code{numpy} and \code{scipy} packages.


\section{Future Development}
\label{future}

Some of the future plans for \code{cluster-lensing} include adding support for different density profiles. Currently only the NFW model is provided, and alternative mass density models would make the package more complete and useful. The first priority will be inclusion of the Einasto profile \citep{Einasto65}, and later possibilities may include the generalized-NFW \citep{Zhao96}. The default cosmology is currently that of \citet{PlanckXVI}, but should be updated to \citet{PlanckXIII_15}, since this is now available as \code{astropy.cosmology.Planck15} (the user can currently specify this cosmology, it is just not the default).

When surface mass density profiles have to be calculated many times for many clusters, as is the case when iterating over parameters in the process of fitting a model, the processing time can become lengthy. This issue is most pronounced for calculation of miscentered profiles, which require the convolution laid out in Equations \ref{sigma_off} and \ref{sigma_RgivenRoff}. One major improvement to \code{cluster-lensing} will be the option to use parallel-processing in these computations. The likely structure of this parallelism will be to divide the halos in a \code{ClusterEnsemble()} catalog object among the parallel threads, which will calculate the profiles for each of their assigned clusters.

All of these future developments are currently listed as issues on the GitHub repository. This GitHub Issue tracker\footnote{\url{https://github.com/jesford/cluster-lensing/issues}} will continue to serve as the central place for listing future improvements and feature requests. Users and potential-users alike are encouraged to submit ideas and requests through that URL.


\section{Summary}
\label{summary}

In this work we presented \code{cluster-lensing}, a pure-\code{Python} package for calculating galaxy cluster profiles and properties. We described and gave worked examples of all the functionality currently available, including mass-concentration and mass-richness scaling relations, and the surface mass density profiles $\Sigma(R)$ and $\Delta\Sigma(R)$, which are relevant for gravitational lensing. The latter density profiles are not cluster-specific, but apply to any mass halo that can be approximated by the NFW prescription. The structure of \code{cluster-lensing} is ideal for calculating properties and profiles for many galaxy clusters at once. This ``composite-halo'' approach \citep[i.e.][]{Ford15}, is especially useful for fitting models to a stacked sample of clusters that span a range of mass and/or redshift.

Compared to existing code, \code{cluster-lensing} stands out by seemingly being the only publicly-available software for calculating miscentered halo profiles. Miscentering is a problem of great relevance for stacked weak lensing studies of galaxy clusters, where halo centers are imperfectly estimated from observational data or simply not well defined (as is the case for individual non-spherical halos -- for example in merging systems). The resulting offsets between the assumed and real centers change the shape of the measured shear or magnification profile and need to be accounted for in the modeling.

\code{cluster-lensing} is released under the MIT license, and archived on Zenodo \citep{clusterlensing}. It being developed in a public repository on GitHub: \url{http://github.com/jesford/cluster-lensing/}. Contributions to the code can be made by submitting a pull request to the repository, and we welcome feedback, suggestions, and feature requests through GitHub issues, or by emailing the author. Full documentation (including much of the content of this paper), as well as installation instructions and examples, are available in the online documentation, at \url{http://jesford.github.io/cluster-lensing/}. If \code{cluster-lensing} is used in a research project, the authors would appreciate citations to the code \citep[i.e.][]{clusterlensing} and this paper.



\section*{Acknowledgements}
The authors are grateful for funding from the Washington Research Foundation Fund for Innovation in Data-Intensive Discovery and the Moore/Sloan Data Science Environments Project at the University of Washington. This project would not have been possible without packages available in Python's open scientific ecosystem, including NumPy \citep{NumPy}, SciPy \citep{SciPy}, Pandas \citep{Pandas}, matplotlib \citep{matplotlib}, IPython \citep{IPython}, AstroPy \citep{astropy13}, Seaborn \citep{Seaborn}, and related tools.

\bibliographystyle{aasjournal}
\bibliography{References}

\end{document}